\documentclass[11pt,twocolumn]{article}
\usepackage{graphicx}
\usepackage{amsmath}
\usepackage{bm}

\newcommand\T{\rule{0pt}{2.6ex}}

\begin{document}

\title{\textsf{\textbf{Taxonomy of Clifford $\mathcal Cl_{3,0}$ subgroups: Choir and band groups}}}
\author{Quirino M. Sugon Jr.* and Daniel J. McNamara
\smallskip\\
\small{Ateneo de Manila University, Department of Physics, Loyola Heights, Quezon City, Philippines 1108}\\
\small{*Also at Manila Observatory, Upper Atmosphere Division, Ateneo de Manila University Campus}\\
\small{e-mail: \texttt{qsugon$@$observatory.ph}}}
\date{\small{\today}}
\maketitle

\section*{}
\small{\textbf{Abstract.}} 
We list the subgroups of the basis set of $\mathcal Cl_{3,0}$  and classify them according to three criteria for construction of universal Clifford algebras: (1) each generator squares to $\pm 1$, (2) the generators within the group anticommute, and (3) the order of the resulting group is $2^{n+1}$, where $n$ is the number of nontrivial generators. Obedient groups we call choirs; disobedient groups, bands.  We classify choirs by modes and bands by rhythms, based on canonical equality.  Each band generator has a transposition (number of other generators it commutes with).  The band's transposition signature is the band's chord.  The sum of transpositions divided by twice the number of generator pair combinations is the band's beat. The band's order deviation is the band's disorder.  For $n\leq 3$, we show that the $\mathcal Cl_{3,0}$ basis set has 21 non-isomorphic subgroups consisting of 9 choirs and 12 bands.
\section{Introduction}

\textbf{a.  Generators of Clifford Algebra.}  Clifford (geometric) algebra is usually presented by taking a set of $n$ unit vectors,
\begin{equation}
\label{eq:B_n1}
\textbf{\textsf B}_{n,1}=\{\mathbf e_1,\mathbf e_2,\ldots,\mathbf e_n\},
\end{equation}
and defining the vector elements to be orthonormal via the following relations\cite{Dimakis_1986p50}:
\begin{eqnarray}
\label{eq:e_j squared is +1}
\mathbf e_j^2 &=& +1,\quad 1\leq j\leq p,\\
\label{eq:e_j squared is -1}
&=& -1,\quad p+1\leq j\leq (p+q=n),\\
\label{eq:e_j e_k is -e_k e_j Cl_pq}
\mathbf e_j\mathbf e_k&=&-\mathbf e_k\mathbf e_j,
\end{eqnarray}
where $j$ and $k$ are nonnegative integers.  The set of all possible products of these $n$ vectors form a group, and from this group we form a group algebra called Clifford (Geometric) Algebra $\mathcal Cl_{p,q}$.  

In the language of combinatorial group theory\cite{MagnusKarassSolitar_1966pp4-7}, we say that $\textbf{\textsf B}_{n,1}$ is the set of generators and Eqs. (\ref{eq:e_j squared is +1}) to (\ref{eq:e_j e_k is -e_k e_j Cl_pq}) are the defining relations.  The word \emph{generator} suggests a mathematical machinery for producing a group, as an electric generator produces electricity.  But the \emph{defining relations} do not provide this machinery as electromagnetic theory do not make an electromagnet.  What we need is a mathematical machinery for listing down all the elements of the geometric group, the group consisting of all basis elements in geometric algebra, with their $+$ and $-$ signs included.

To answer this need, we shall use a set algebra formalism with two operations: multiplication (juxtaposition) and set union ($\cup$).  This algebra is already known, but remains unused for listing group elements.  

In group theory, the product of two subgroups $\textbf{\textsf F}$ and $\textbf{\textsf H}$ of group \textbf{\textsf G} is defined as
\begin{equation}
\textbf{\textsf F}\textbf{\textsf H}=\{\hat f\hat h\ |\ \textrm{for all} \ \hat f\in\textbf{\textsf F} ,\hat h\in\textbf{\textsf H}\}.
\end{equation}
Scherphuis\cite{Scherphuis_2008} commented that this definition need not be restricted to subgroups: \textbf{\textsf F} and \textbf{\textsf H} need only be subsets of \textbf{\textsf G}.  If we adopt this view, then we see, for example, that the distributivity property holds:
\begin{eqnarray}
\label{eq:H_1 over H_2 U H_3 distributivity intro}
\textbf{\textsf H}_1(\textbf{\textsf H}_2\cup\textbf{\textsf H}_3)&=&(\textbf{\textsf H}_1\textbf{\textsf H}_2)\cup(\textbf{\textsf H}_1\textbf{\textsf H}_3),\\
\label{eq:H_2 U H_3 over H_1 distributivity intro}
(\textbf{\textsf H}_2\cup\textbf{\textsf H}_3)\textbf{\textsf H}_1&=&(\textbf{\textsf H}_2\textbf{\textsf H}_1)\cup(\textbf{\textsf H}_3\textbf{\textsf H}_1),
\end{eqnarray}
where $\textbf{\textsf H}_1,\textbf{\textsf H}_2,\textbf{\textsf H}_3\in\textbf{\textsf G}$.  Thus, it is indeed possible to create a set algebra over set union.

Johnson\cite{Johnson_2000p99} may have some idea of this set multiplication when he wrote the Clifford algebraic set equation
\begin{equation}
\label{eq:set 1 e_2 set 1 e_1 is set 1 e_1 e_2 e_1e_2}
\{1,\mathbf e_2\}\otimes\{1,\mathbf e_1\}=\{1,\mathbf e_1,\mathbf e_2,\mathbf e_1\mathbf e_2\}.
\end{equation}
But his $\otimes$ operator appears to mean differently: find all the linearly independent products of the elements of the sets $\{1,\mathbf e_1\}$ and $\{1,\mathbf e_2\}$.  Clearly, this operator does not always satisfy Eq.~(\ref{eq:H_1 over H_2 U H_3 distributivity intro}) (it would if we choose $\mathbf e_2\mathbf e_1$ instead of $\mathbf e_1\mathbf e_2$, but this decision is arbitrary).

Using the set algebra we propose, we can show, for example, that the set $\textbf{\textsf C}_{2,0}$ of all basis elements in $\mathcal Cl_{2,0}$ can be expressed in two ways:
\begin{eqnarray}
\label{eq:C_20 as e_1 e_2}
\textbf{\textsf C}_{2,0}&=&\{1,-1\}\{1,\mathbf e_1\}\{1,\mathbf e_2\},\\
\label{eq:C_20 as e_1 i}
&=&\{1,-1\}\{1,\mathbf e_1\}\{1,\hat\imath\},
\end{eqnarray}
where $\hat\imath =\mathbf e_1\mathbf e_2$.  Notice the similarity of Eq.~(\ref{eq:C_20 as e_1 e_2}) with Eq.~(\ref{eq:set 1 e_2 set 1 e_1 is set 1 e_1 e_2 e_1e_2}), save for the presence of $\{1,-1\}$, which prevents us from worrying about the order of the factors.  The set product form in Eq.~(\ref{eq:C_20 as e_1 e_2}) or (\ref{eq:C_20 as e_1 i}) is the mathematical machinery we need for listing the elements of a geometric group like $\textbf{\textsf C}_{2,0}$.  From this group we construct a geometric group algebra.

\textbf{b.  Signature of Clifford Algebra.}  But what geometric algebra do we choose from all the possible dialects of $\mathcal Cl_{p,q}$?  Is $\mathbf e_j^2=+1$ or $-1$ or both?  

The axiom $\mathbf e_j^2=-1$ is a good choice if we wish to illustrate the natural continuity of Clifford algebra $\mathcal Cl_{0\!,\,n}$ with Hamilton's quaternions\cite{Gallier_2008p15}.  In fact, this is what Clifford chose in 1878.  But in 1882, he changed his mind and wrote $\mathbf e_j^2=1$\cite{Lounesto_1993p216}.  What made Clifford change his mind, we can only guess.  Perhaps, it is the notion that the square of a vector must be its magnitude squared; quaternions fail this test.  This is what Cayley in 1871 pointed out to the quaternion defender Tait\cite{Pritchard_1998p236}.  And this is also what Gibbs in 1870's thought before his word war with Tait\cite{Pritchard_1998p238}.  So in deference to the opinions of Clifford, Cayley, and Gibbs, we shall adopt the axiom $\mathbf e_j^2=1$.  Thus, the algebra that we shall use is not $\mathcal Cl_{0,n}$ but $\mathcal Cl_{n\!,\,0}$.

By restricting ourselves to positive signatured Clifford algebras, we claim that we can still obtain other Clifford algebras.  In $\mathcal Cl_{3,0}$, for example, we can construct basis sets isomorphic to that of $\mathcal Cl_{1,2}$, $\mathcal Cl_{2,0}$, $\mathcal Cl_{1,1}$, $\mathcal Cl_{0,2}$, $\mathcal Cl_{1,0}$, $\mathcal Cl_{0,1}$, and $\mathcal Cl_{0,0}$.  These basis sets we shall call choirs.  Basis sets that do not construct Clifford algebras we shall call bands.  We shall see that from the basis set of $\mathcal Cl_{3,0}$, we can form 21 non-isomorphic subroups consistings of 9 choirs and 12 bands.

\textbf{c.  Outline.}  We shall divide the body of the paper into five sections.  The first section is Introduction.  In the second section, we shall discuss vector products via logic gate operations and Walsh functions, and use these to prove the group properties of the basis set of Clifford algebra $\mathcal Cl_{n,0}$.  In the third section, we shall list the axioms of set algebra with two operations: juxtaposition multiplication and set union.  We shall define and investigate geometric groups, which are similar to Clifford basis groups, except that their generators are allowed to commute and their squares are allowed to be negative.  In the fourth section, we shall list the geometric subgroups of the basis set of $\mathcal Cl_{3,0}$ and classify them into choirs and bands.  We shall arrange choirs by modes and bands by rhythms.  We shall distinguish bands by their Clifford signature, (dis)order of elements, and number of commuting generators.  The last section is Conclusions.

\section{Clifford Basis Group}

\subsection{Axioms}

Let us define 
\begin{eqnarray}
\label{eq:G_31 start}
\textbf{\textsf G}_{n,1}=\{\pm\mathbf e_1,\pm\mathbf e_2,\ldots,\pm\mathbf e_j,\ldots,\pm\mathbf e_n\}.
\end{eqnarray}
The quantity $\mathbf e_j$ where $j\in\{1,2,\ldots,n\}$ is called a vector or ray and $-\mathbf e_j$ is the vector opposite to it.  

Let us assume that the six vectors in the set $\textbf{\textsf G}_{n,1}$ satisfy the three axioms for the geometric product (denoted by juxtaposition multiplication):
\bigskip

\textbf{a.  Associativity.}  For all $j,k,\ell\in\{1,2,3\}$,
\begin{equation}
\label{eq:e_j e_k e_l associativity}
(\mathbf e_j\mathbf e_k)\mathbf e_\ell=\mathbf e_j(\mathbf e_k\mathbf e_\ell).
\end{equation}
That is, regrouping the factors do not affect the product.

\textbf{b.  Orthogonality.}  For $j,k\in\{1,2,3\}$,
\begin{equation}
\label{eq:e_j e_k is -e_k e_j}
\mathbf e_j\mathbf e_k=-\mathbf e_k\mathbf e_j.
\end{equation}
Geometrically, $\mathbf e_j\mathbf e_k$ is the oriented area defined by the vectors (rays) $\mathbf e_j$ and $\mathbf e_k$ connected head-to-tail according to the order of the factors.  Thus, $\mathbf e_k\mathbf e_j$ is the same area but oriented opposite.  This explains the negative sign.

\textbf{b.  Norm.}  For $j\in\{1,2,3\}$,
\begin{equation}
\label{eq:e_j squared is e_k squared is 1}
\mathbf e_j^2=1.\\
\end{equation}
Geometrically, Eq.~(\ref{eq:e_j squared is e_k squared is 1}) states that the length of $\mathbf e_j$ is one unit.

\subsection{Theorems}

From these three axioms we can prove three theorems:
\bigskip

\textbf{a.  Power.} The raising of a vector $\mathbf e_j$ for $j\in\{1,2,3\}$ to a nonnegative integer power $b_j$ results to either the real number $1$ or the vector itself: 
\begin{eqnarray}
\label{eq:e_j to b_j is 1}
\mathbf e_j^{b_j}=&\!\!1,\qquad b_j\equiv&\!\!\! 0\ \textrm{mod}\ 2,\\
=&\!\!\mathbf e_j,\qquad b_j\equiv&\!\!\! 1\ \textrm{mod}\ 2.
\end{eqnarray}
That is, if the power is even, $\mathbf e_j^{b_j}=1$; if odd, $\mathbf e_j^{b_j}=\mathbf e_j$.

One corollary to this theorem is that the inverse of $\mathbf e_j^{b_j}$ is itself:
\begin{equation}
\label{eq:e_j to b_j inverse}
(\mathbf e_j^{b_j})^{-1}=\mathbf e_j^{b_j}=\mathbf e_j^{-b_j}.
\end{equation}
That is,
\begin{equation}
\label{eq:e_j to b_j inverse e_j to b_j}
(\mathbf e_j^{b_j})^{-1}\mathbf e_j^{b_j}=\mathbf e_j^{b_j}\mathbf e_j^{b_j}=\mathbf e_j^{2b_j}=1,
\end{equation}
because $2b_j\equiv 0\ \textrm{mod}\ 2$.

\textbf{b.  XOR.}  The product of $\mathbf e_j^{b_j}$ and $\mathbf e_j^{b_j'}$ is
\begin{equation}
\label{eq:e_j to b_j e_j to b_j'}
\mathbf e_j^{b_j}\mathbf e_j^{b_j'}=\mathbf e_j^{b_j+b_j'}.
\end{equation}
There are four possibilities:
\begin{eqnarray}
\label{eq:e_j to b_j+b_j' case 00}
\mathbf e_j^{b_j+b_j'}&=\mathbf e_j^0,\quad& b_j=0,\ b_j'=0;\\
\label{eq:e_j to b_j+b_j' case 01}
&=\mathbf e_j^1,\quad& b_j=0,\ b_j'=1;\\
\label{eq:e_j to b_j+b_j' case 10}
&=\mathbf e_j^1,\quad& b_j=1,\ b_j'=0;\\
\label{eq:e_j to b_j+b_j' case 11}
&=\mathbf e_j^0,\quad& b_j=1,\ b_j'=1.
\end{eqnarray}
Thus, in terms of the XOR gate, we write
\begin{equation}
\label{eq:e_j to b_j XOR b_j'}
\mathbf e_j^{b_j+b_j'}=\mathbf e_j^{b_j\ {\textrm{\tiny XOR}}\ b_j'}.
\end{equation}

\textbf{c.  AND.}  The product of $\mathbf e_j^{b_j}$ and $\mathbf e_{k}^{b_k}$  for $j\neq k$ is governed by the orthogonality relation in Eq.~(\ref{eq:e_j e_k is -e_k e_j}).  This product has four possibilities:
\begin{eqnarray}
\label{eq:e_j to b_j e_k to b_k case 00}
\mathbf e_j^{b_j}\mathbf e_k^{b_k}=&\mathbf e_k^{b_k}\mathbf e_j^{b_j},\quad& b_j=0,\ b_{k}=0;\\
\label{eq:e_j to b_j e_k to b_k case 01}
=&\mathbf e_k^{b_k}\mathbf e_j^{b_j},\quad& b_k=0,\ b_{k}=1;\\
\label{eq:e_j to b_j e_k to b_k case 10}
=&\mathbf e_k^{b_k}\mathbf e_j^{b_j},\quad& b_k=1,\ b_{k}=0;\\
\label{eq:e_j to b_j e_k to b_k case 11}
=&-\mathbf e_k^{b_k}\mathbf e_j^{b_j},\quad& b_k=1,\ b_{k}=1.
\end{eqnarray}
In terms of the AND gate, we write
\begin{equation}
\label{eq:e_j to b_j e_k to b_k is -1 AND flip}
\mathbf e_j^{b_j}\mathbf e_k^{b_k}=(-1)^{b_j\ {\textrm{\tiny AND}}\ b_k}\,\mathbf e_k^{b_k}\mathbf e_j^{b_j}=(-1)^{b_jb_k}\,\mathbf e_k^{b_k}\mathbf e_j^{b_j}.
\end{equation}

\subsection{Group Properties}

Let $\textbf{\textsf G}_n$ be a set whose elements are of the form
\begin{equation}
\label{eq:g}
\hat g=s_g\mathbf e_1^{b_1}\mathbf e_2^{b_2}\cdots\mathbf e_k^{b_k}\cdots\mathbf e_n^{b_n},
\end{equation}
where $s_g=\pm 1$ is the sign of $\hat g$ and the bit power $b_k\in\{0,1\}$ for $k\in\{1,2,\ldots,n\}$.  Our aim is to show that $\textbf{\textsf G}_n$ satisfies the four group properties under geometric product: closure, associativity, identity, and inverse\cite{ChenPingWang_2002p4}.  We shall call $\textbf{\textsf G}_n$ as the Clifford basis group with $n$ vector generators.
\bigskip

\textbf{a.  Closure.}  Let $\hat g$ and $\hat g'$ be two elements of $\textbf{\textsf G}_n$.  Their product is
\begin{equation}
\label{eq:gg'}
\hat g\hat g'=s_{\hat g}s_{\hat g'}\mathbf e_1^{b_1}\mathbf e_2^{b_2}\cdots\mathbf e_j^{b_j}\cdots\mathbf e_n^{b_n}\mathbf e_1^{b_1'}\mathbf e_2^{b_2'}\cdots\mathbf e_k^{b_k'}\cdots\mathbf e_n^{b_n'}.
\end{equation}
Using the theorems in Eqs.~(\ref{eq:e_j to b_j e_j to b_j'}) and (\ref{eq:e_j to b_j e_k to b_k is -1 AND flip}), Eq.~(\ref{eq:gg'}) becomes
\begin{equation}
\label{eq:gg' simplified}
\hat g\hat g'=W_{\hat g\hat g'}s_{\hat g}s_{\hat g'}\mathbf e_1^{b_1+b'_1}\mathbf e_2^{b_2+b'_2}\cdots\mathbf e_n^{b_n+b'_n},
\end{equation}
where $W_{\hat g\hat g'}$ is a Walsh function,
\begin{eqnarray}
\label{eq:W_gg'}
W_{\hat g\hat g'}&=&(-1)^{M_{\hat g\hat g'}},\\
\label{eq:M_gg'}
M_{\hat g\hat g'}&=&\sum_{k=1}^{n-1}b_k'\sum_{j\,=\,k+1}^nb_j.
\end{eqnarray}
Because Eq.~(\ref{eq:e_j to b_j XOR b_j'}) is true and $W_{\hat g\hat g'}=\pm 1$, then $\hat g\hat g'$ in Eq.~(\ref{eq:gg'}) is an element of $\textbf{\textsf G}_n$.  Thus, $\textbf{\textsf G}_n$ is closed under juxtaposition multiplication.  

The $\mathcal Cl_{n,0}$ Walsh function in Eq.~(\ref{eq:W_gg'}) is given in Biasbas\cite{Biasbas_2005p7}, while that for $\mathcal Cl_{0,n}$ was derived by Vahlen (1897)\cite{Vahlen_1897p217} and rederived by Hagmark and Lounesto\cite{HagmarkLounesto_1986p536}.  The general Walsh function for $\mathcal Cl_{p,q}$ is derived by Brauer and Weyl (1935) \cite{BrauerWeyl_1935p536}.

\textbf{b.  Associativity.}  For $\hat g_1,\hat g_2,\hat g_3\in\textbf{\textsf G}_n$,
\begin{equation}
\label{eq:g_1 g_2 g_3 associativity}
\hat g_1(\hat g_2\hat g_3)=(\hat g_1\hat g_2)\hat g_3,
\end{equation}
because products of the basis vectors in the set $\textbf{\textsf B}_{n,1}$ in Eq.~(\ref{eq:B_n1}) are associative.

\textbf{c.  Identity.}  The unit real number $1$ is the unit element of $\textbf{\textsf G}_n$, because
\begin{equation}
\label{eq:identity is 1 in G_n}
1=\mathbf e_1^0\mathbf e_2^0\cdots\mathbf e_n^0,
\end{equation}
which is of the form of an element of $\textbf{\textsf G}_n$.  Hence,
\begin{equation}
\label{eq:1g is g1 is g}
1\hat g=\hat g1=\hat g.
\end{equation}

\textbf{d.  Inverse.}  For every $\hat g\in\textbf{\textsf G}_n$, there exists $\hat g^{-1}\in\textbf{\textsf G}_n$ such that
\begin{equation}
\label{eq:gg inverse is g inverse g is 1}
\hat g\hat g^{-1}=\hat g^{-1}\hat g=1.
\end{equation}
From the definition of $\hat g$ in Eq.~(\ref{eq:g}) and the identity $\mathbf e_k^{2b_k}=1$ from Eq.~(\ref{eq:e_j to b_j is 1}), we can immediately see that 
\begin{equation}
\label{eq:g inverse}
\hat g^{-1}=s_{\hat g}\mathbf e_n^{b_n}\cdots\mathbf e_k^{b_k}\ldots\mathbf e_2^{b_2}\mathbf e_1^{b_1}.
\end{equation}
Using the theorem in Eq.~(\ref{eq:e_j to b_j e_k to b_k is -1 AND flip}), Eq.~(\ref{eq:g inverse}) may be written as
\begin{equation}
\label{eq:g inverse standard form}
\hat g^{-1}=s_{\hat g}W_{\hat g^{-1}}\mathbf e_1^{b_1}\mathbf e_2^{b_2}\cdots\mathbf e_k^{b_k}\cdots\mathbf e_n^{b_n},
\end{equation}
where
\begin{eqnarray}
\label{eq:W_g inverse}
W_{\hat g^{-1}}&=&(-1)^{M_{\hat g^{-1}}},\\
\label{eq:M_gg inverese}
M_{\hat g^{-1}}&=&\sum_{k=1}^{n-1}b_k\sum_{j\,=\,k+1}^nb_j.
\end{eqnarray}
Because $W_{\hat g^{-1}}=\pm 1$, then $g^{-1}\in\textbf{\textsf G}_n$.

\section{Geometric Group}

\subsection{Set Axioms}

Let $\textbf{\textsf F},\textbf{\textsf H}\subseteq\textbf{\textsf G}$.  We define the product of the sets $\textbf{\textsf F}$ and $\textbf{\textsf H}$ as follows:
\begin{eqnarray}
\label{eq:FH}
\textbf{\textsf F}\textbf{\textsf H}=\{\hat f\hat h\ |\ \hat f\in\textbf{\textsf F},\hat h\in\textbf{\textsf H}\}.
\end{eqnarray}
From this definition we can construct an algebra for the set $\textbf{\textsf G}$ with respect to juxtaposition multiplication and set union $\cup$:
\bigskip

\textbf{a.  Associativity.}  If, $\textbf{\textsf H}_1,\textbf{\textsf H}_2,\textbf{\textsf H}_3\in\textbf{\textsf G}$, then
\begin{eqnarray}
\label{eq:H_1 H_2 H_3 associativity}
(\textbf{\textsf H}_1\textbf{\textsf H}_2)\textbf{\textsf H}_3&=&\textbf{\textsf H}_1(\textbf{\textsf H}_2\textbf{\textsf H}_3),\\
(\textbf{\textsf H}_1\cup\textbf{\textsf H}_2)\cup\textbf{\textsf H}_3&=&\textbf{\textsf H}_1\cup(\textbf{\textsf H}_2\cup\textbf{\textsf H}_3).
\end{eqnarray}

\textbf{b.  Commutativity.}  If $\textbf{\textsf H}_1,\textbf{\textsf H}_2\in\textbf{\textsf G}$, then
\begin{equation}
\label{eq:H_1 U H_2 commutativity}
\textbf{\textsf H}_1\cup\textbf{\textsf H}_2=\textbf{\textsf H}_2\cup\textbf{\textsf H}_1.
\end{equation}

\textbf{c.  Identity.}  There exists an identity $\{1\}\in\textbf{\textsf G}_3$ and an empty set $\{\}\in\textbf{\textsf G}$, such that for all $\textbf{\textsf H}\subseteq\textbf{\textsf G}$,
\begin{eqnarray}
\label{eq:H 1 is H}
\textbf{\textsf H}\{1\}=\{1\}\textbf{\textsf H}&=&\textbf{\textsf H},\\
\label{eq:H U null is H}
\textbf{\textsf H}\cup\{\}=\{\}\cup\textbf{\textsf H}&=&\textbf{\textsf H}.
\end{eqnarray}

\textbf{d.  Distributivity.}  If, $\textbf{\textsf H}_1,\textbf{\textsf H}_2,\textbf{\textsf H}_3\in\textbf{\textsf G}$, then
\begin{eqnarray}
\label{eq:H_1 over H_2 U H_3}
\textbf{\textsf H}_1(\textbf{\textsf H}_2\cup\textbf{\textsf H}_3)&=&\textbf{\textsf H}_1\textbf{\textsf H}_2\cup\textbf{\textsf H}_1\textbf{\textsf H}_3,\\
\label{eq:H_2 U H_3 over H_1}
(\textbf{\textsf H}_2\cup\textbf{\textsf H}_3)\textbf{\textsf H}_1&=&\textbf{\textsf H}_2\textbf{\textsf H}_1\cup\textbf{\textsf H}_3\textbf{\textsf H}_1.
\end{eqnarray}

\textbf{e.  Inverse.}  If $\hat g\in\textbf{\textsf G}$, then there exists a $\hat g^{-1}\in\textbf{\textsf G}_3$ such that
\begin{equation}
\label{eq:g g inverse sets}
\hat g\hat g^{-1}=\hat g^{-1}\hat g=1,
\end{equation}
which is a group property.  Sets containing more than one element has no multiplicative inverse.  There is also no set that when united to a nonempty set gives the empty set $\{\}$.  

\subsection{Set Theorems}

From these five set properties, we can derive two theorems:
\bigskip

\textbf{a. Subset-Group Product.}  If $\textbf{\textsf G}$ is a group and  $\textbf{\textsf H}\subseteq\textbf{\textsf G}$, then
\begin{equation}
\label{eq:HG is G}
\textbf{\textsf H}\textbf{\textsf G}=\textbf{\textsf G}\textbf{\textsf H}=\textbf{\textsf G}.
\end{equation}
That is, the product of a group and its subset is the group itself.

\emph{Proof}. Let us only prove $\textbf{\textsf H}\textbf{\textsf G}=\textbf{\textsf G}$; the proof for $\textbf{\textsf G}\textbf{\textsf H}=\textbf{\textsf G}$ is similar.

Let $\hat h\in\textbf{\textsf H}$.  Because $\textbf{\textsf H}\subseteq\textbf{\textsf G}$, then $\hat h\in\textbf{\textsf G}$.  Since $\textbf{\textsf G}$ is a group, then the left coset of $\textbf{\textsf G}$ is $\textbf{\textsf G}$ itself:
\begin{equation}
\label{eq:hG is G}
\hat h\textbf{\textsf G}=\textbf{\textsf G}.
\end{equation}
Thus, if $\textbf{\textsf H}$ is a set with $m$ elements,
\begin{equation}
\label{eq:set H with m elements}
\textbf{\textsf H}=\{\hat h_1,\hat h_2,\ldots,\hat h_m\},
\end{equation}
then
\begin{eqnarray}
\label{eq:HG is G expanded}
\textbf{\textsf H}\textbf{\textsf G}&=&\hat h_1\textbf{\textsf G}\cup\hat h_2\textbf{\textsf G}\cup\cdots\cup\hat h_m\textbf{\textsf G}\nonumber\\
&=&\textbf{\textsf G}\cup\textbf{\textsf G}\cup\cdots\cup\textbf{\textsf G}\nonumber\\
&=&\textbf{\textsf G},
\end{eqnarray}
which is what we wish to prove. {\tiny QED}

One corollary to the theorem in Eq.~(\ref{eq:HG is G}) is that $\textbf{\textsf G}$ is \emph{idempotent}, i.e., its square is itself:
\begin{equation}
\label{eq:G squared is G}
\textbf{\textsf G}^2=\textbf{\textsf G}.
\end{equation}
This equation is an alternative statement for the closure property of $\textbf{\textsf G}$.  

The set $\textbf{\textsf R}$ of real numbers is also idempotent, for $\textbf{\textsf R}^2=\textbf{\textsf R}$.  And so is the set $\{\pm 1\}\equiv\{1,-1\}$:
\begin{equation}
\label{eq:set pm 1 squared  is set pm 1}
\{\pm 1\}^2=\{1,-1\}\{1,-1\}=\{1,-1\}=\{\pm 1\}.
\end{equation}
This group is isomorphic to $\textbf{\textsf Z}_2=\{0,1\}$  under addition mod 2.

\textbf{b.  Commutation with $\{\pm 1\}$.}  If $\textbf{\textsf G}$ is a group and $\hat g\in\textbf{\textsf G}$, then
\begin{equation}
\label{eq:set pm 1 set 1 g commute}
\{\pm 1\}\{1,\hat g\}=\{1,\hat g\}\{\pm 1\},
\end{equation}

\emph{Proof.}  We simply show that the expansions of the left and right sides of Eq.~(\ref{eq:set pm 1 set 1 g commute}) are the same:
\begin{eqnarray}
\label{eq:set pm 1 set 1 g commute expand}
\{1,-1\}\{1,\hat g\}&=&\{1,-1,\hat g,-\hat g\}\nonumber\\
&=&\{\pm 1,\pm g\}=\{1,\hat g\}\{1,-1\}.
\end{eqnarray}

\subsection{Generators of a Geometric Group}

Let $\textbf{\textsf G}$ be a group and let $\textbf{\textsf H}\subseteq\textbf{\textsf G}$ defined in Eq.~(\ref{eq:set H with m elements}) satisfy two conditions:
\begin{itemize}
\item For all $\hat h_j\in\textbf{\textsf H}$,
\begin{equation}
\label{eq:h_j squared is pm 1}
\hat h_j^2=\pm 1.
\end{equation}

\item For all $\hat h_j,\hat h_k\in\textbf{\textsf H}$, $\hat h_j$ and $\hat h_k$ either commute or anticommute:
\begin{equation}
\label{eq:h_j h_k is pm h_k h_j}
\hat h_j\hat h_k=\pm\hat h_k\hat h_j.
\end{equation}
\end{itemize}
We claim that the set
\begin{equation}
\label{eq:G_H}
\textbf{\textsf G}_{\textbf{\textsf H}}=\{\pm 1\}\{1,h_1\}\{1,h_2\}\cdots\{1,h_m\}
\end{equation}
generated from $\textbf{\textsf H}$ is a group.  We call $\textbf{\textsf G}_{\textbf{\textsf H}}$ as the geometric group generated by the set $\textbf{\textsf H}$.

\emph{Proof.}  We verify that $\textbf{\textsf G}_{\textbf{\textsf H}}$ satisfies the closure, associativity, identity, and inverse properties of a group:
\bigskip

\textbf{a.  Closure.}  To prove that $\textbf{\textsf G}_{\textbf{\textsf H}}$ is closed, we use the definition in Eq.~(\ref{eq:G squared is G}):
\begin{equation}
\label{eq:G_H squared is G_H}
\textbf{\textsf G}_{\textbf{\textsf H}}^2=\textbf{\textsf G}_{\textbf{\textsf H}}.
\end{equation}

The left side of Eq.~(\ref{eq:G_H squared is G_H}) may be expanded as
\begin{eqnarray}
\label{eq:G_H squared expand}
\textbf{\textsf G}_{\textbf{\textsf H}}^2&=&\{\pm 1\}\{1,h_1\}\{1,h_2\}\cdots\{1,h_m\}\cdot\nonumber\\
& &\{\pm 1\}\{1,h_1\}\{1,h_2\}\cdots\{1,h_m\},
\end{eqnarray}
where the centered dot $(\cdot)$ means that the other set factors are below the line.  Using the theorems in Eqs.~(\ref{eq:set pm 1 set 1 g commute}) and (\ref{eq:set pm 1 squared  is set pm 1}), we may move the middle $\{\pm 1\}$ to the place beside the leftmost $\{\pm 1\}$ to arrive at
\begin{eqnarray}
\label{eq:G_H squared expand middle set pm 1 removed}
\textbf{\textsf G}_{\textbf{\textsf H}}^2&=&\{\pm 1\}\{1,h_1\}\{1,h_2\}\cdots\{1,h_m\}\cdot\nonumber\\
& &\{1,h_1\}\{1,h_2\}\cdots\{1,h_m\},
\end{eqnarray}
because $\{\pm 1\}$ is idempotent.

Now, we move $\{\pm 1\}$ again near the middle part.  Because $\hat h_1$ commutes or anticommutes with $\hat h_m$ by Eq.~(\ref{eq:h_j h_k is pm h_k h_j}), then 
\begin{eqnarray}
\label{eq:set pm 1 set 1 h_m set 1 h_1 permute}
\{\pm 1\}\{1,\hat h_m\}\{1,\hat h_1\}&=&\{\pm 1,\pm\hat h_m,\pm\hat h_1,\pm\hat h_m\hat h_1\}\nonumber\\
&=&\{\pm 1,\pm\hat h_m,\pm\hat h_1,\pm\hat h_1\hat h_m\}\nonumber\\
&=&\{\pm 1\}\{1,\hat h_1\}\{1,\hat h_m\}.
\end{eqnarray}
Then we move $\{\pm 1\}$ again back to the leftmost side, so that Eq.~(\ref{eq:G_H squared expand middle set pm 1 removed}) becomes
\begin{eqnarray}
\label{eq:G_H squared expand middle set pm 1 permuted factors}
\textbf{\textsf G}_{\textbf{\textsf H}}^2&=&\{\pm 1\}\{1,h_1\}\{1,h_2\}\cdots\{1,h_1\}\cdots\nonumber\\
& &\{1,h_m\}\{1,h_2\}\cdots\{1,h_m\}.
\end{eqnarray}

In general, we have the following \emph{lemma}: if a set is of the form given in Eq.~(\ref{eq:G_H}) and the condition in Eq.~(\ref{eq:h_j h_k is pm h_k h_j}) holds, the order of the set factors does not matter. 

Applying this \emph{lemma} for $\textbf{\textsf G}_{\textbf{\textsf H}}^2$ in Eq.~(\ref{eq:G_H squared expand middle set pm 1 removed}), we get
\begin{equation}
\label{eq:G_H squared paired factors}
\textbf{\textsf G}_{\textbf{\textsf H}}^2=\{\pm 1\}\{1,h_1\}^2\{1,h_2\}^2\cdots\{1,h_m\}^2.
\end{equation}
Because $\hat h_j^2=\pm 1$ by Eq.~(\ref{eq:h_j squared is pm 1}), then
\begin{equation}
\label{eq:set pm 1 set 1 h_k squared is set pm 1 set 1 h_k}
\{\pm 1\}\{1,\hat h_k\}^2=\{\pm 1,\pm\hat h_k\}=\{\pm 1\}\{1,\hat h_k\},
\end{equation}
so that Eq.~(\ref{eq:G_H squared paired factors}) becomes
\begin{equation}
\label{eq:G_H squared is G_H end}
\textbf{\textsf G}_{\textbf{\textsf H}}^2=\{\pm 1\}\{1,h_1\}\{1,h_2\}\cdots\{1,h_m\}=\textbf{\textsf G}_{\textbf{\textsf H}},
\end{equation}
which is what we wish to prove.  {\tiny QED}

\textbf{b.  Associativity.}  We may rewrite the group $\textbf{\textsf G}_{\textbf{\textsf H}}$ in Eq.~(\ref{eq:G_H}) as
\begin{equation}
\label{eq:G_H bit form}
\textbf{\textsf G}_{\textbf{\textsf H}}=\{\pm 1\}\{\hat h_1^0,\hat h_1\}\{\hat h_2^0,\hat h_2\}\cdots\{\hat h_m^0,\hat h_m\}.
\end{equation}
Thus, every $\hat A\in\textbf{\textsf G}_{\textbf{\textsf H}}$ is of the form
\begin{equation}
\label{eq:A element of G_H}
\hat A=s_{\hat A}\hat h_1^{a_1}\hat h_2^{a_2}\cdots\hat h_j^{a_j}\cdots\hat h_m^{a_m},
\end{equation}
where $s_{\hat A}=\pm 1$ is the sign of $\hat A$ and $a_j\in\{0,1\}$ is the bit power of generator $\hat h_j$.

If $\hat B,\hat C\in\textbf{\textsf G}_{\textbf{\textsf H}}$, then
\begin{equation}
\hat A(\hat B\hat C)=(\hat A\hat B)\hat C,
\end{equation}
because the factors of $\hat A$, $\hat B$, and $\hat C$ are elements of the group $\textbf{\textsf G}$, and the products of the elements in a group are associative. {\tiny QED}

\textbf{c.  Identity}  The identity element of $\textbf{\textsf G}_{\textbf{\textsf H}}$ is the real number $1$:
\begin{equation}
\label{eq:Identity of G_H is 1}
\hat h_1^0\hat h_2^0\cdots\hat h_m^0=1,
\end{equation}
so that for all $\hat A\in\textbf{\textsf G}_{\textbf{\textsf H}}$,
\begin{equation}
\label{eq:A1 is 1A in G_H}
\hat A1=1\hat A=\hat A.
\end{equation}

\textbf{d.  Inverse}  From the conditions in Eqs.~(\ref{eq:h_j squared is pm 1}) and (\ref{eq:h_j h_k is pm h_k h_j}), we see that the inverse of $\hat A\in\textbf{\textsf G}_{\textbf{\textsf H}}$ is either $\hat A$ itself or $-\hat A\in\textbf{\textsf G}_{\textbf{\textsf H}}$, so that
\begin{equation}
\label{eq:A A inverse is A inverse A for G_H}
\hat A\hat A^{-1}=\hat A^{-1}\hat A.
\end{equation}

\section{Groups and Subgroups}

The Clifford basis group $\textbf{\textsf C}_{p,q}$ consists of $n=p+q$ anticommuting generators, with $p$ generators that square to $+1$ and $q$ generators that square to $-1$.  This group $\textbf{\textsf C}_{p,q}$ forms the basis for the Clifford group algebra $\mathcal Cl_{p,q}$.

Our aim in this section is to describe the Clifford (Pauli) basis group $\textbf{\textsf C}_{3,0}$ and enumerate its geometric subgroups in a hierarchical order for $n\leq 3$.

\subsection{Subscript Notation}

We shall use the following notations for the following geometric groups with one nontrivial generator ($\neq-1$):
\begin{eqnarray}
\label{eq:E_a definition}
\textbf{\textsf E}_a&=&\{\pm 1\}\{1,\mathbf e_a\},\\
\label{eq:E_ab definition}
\textbf{\textsf E}_{ab}&=&\{\pm 1\}\{1,\mathbf e_a\mathbf e_b\}\equiv\{\pm 1\}\{1,\hat{\mathbf e}_{ab}\},\\
\label{eq:E_abc definition}
\textbf{\textsf E}_{abc}&=&\{\pm 1\}\{1,\mathbf e_a\mathbf e_b\mathbf e_c\}\equiv\{\pm 1\}\{1,\hat{\mathbf e}_{abc}\},
\end{eqnarray}
where $a,b,c\in\{1,2,3\}$ are distinct integers.  Note that the product of these groups is also a geometric group.  For example, the Clifford basis group $\textbf{\textsf C}_{3,0}$ may be represented as
\begin{eqnarray}
\label{eq:E_a E_b E_c example}
\textbf{\textsf E}_a\textbf{\textsf E}_{b}\textbf{\textsf E}_{c}&=&\{\pm 1\}^3\{1,\mathbf e_a\}\{1,\hat{\mathbf e}_{b}\}\{1,\hat{\mathbf e}_{c}\}\nonumber\\
&=&\{\pm 1\}\{1,\mathbf e_a,\mathbf e_b,\hat{\mathbf e}_{ab}\}\{1,\mathbf e_c\}\nonumber\\
&=&\{\pm\}\{1,\mathbf e_a,\mathbf e_b,\mathbf e_c,\hat{\mathbf e}_{ab},\hat{\mathbf e}_{ac},\hat{\mathbf e}_{bc},\hat{\mathbf e}_{abc}\}\nonumber\\
&=&\textbf{\textsf C}_{3,0}.
\end{eqnarray} 

\subsection{Four Group Relations}

There are four relations that characterize groups:
\bigskip

\textbf{a.  Isomorphic.}  Two geometric groups are isomorphic ($\cong$) if there exists a one-to-one mapping between their generators, which preserve the relationship between the generators (commutation and anticommutation) and the square of each generator ($\pm 1$).  For example,
\begin{eqnarray}
\label{eq:E_a isomorphic C_10 example}
\textbf{\textsf E}_a&\cong&\textbf{\textsf C}_{1\!,\,0},\\
\label{eq:E_ab isomorphic C_01 example}
\textbf{\textsf E}_{ab}&\cong&\textbf{\textsf C}_{0,1}.
\end{eqnarray}

\textbf{b.  Similar.}  Two geometric groups are similar ($\approx$) if they have the same generators, except for the labelling of the generator subscripts.  For example,
\begin{equation}
\label{eq:E_abc permute}
\textbf{\textsf E}_{abc}\approx\textbf{\textsf E}_{acb}\approx\textbf{\textsf E}_{bac},
\end{equation}
but
\begin{equation}
\label{eq:E_ab E_bc not similar E_ab E_cd}
\textbf{\textsf E}_{ab}\textbf{\textsf E}_{bc}\approx\!\!\!\!\!\!/\ \ \textbf{\textsf E}_{ab}\textbf{\textsf E}_{cd},
\end{equation}
because the mapping $b:\rightarrow\{b,d\}$ is not one-to-one.  Note that all similar groups are isomorphic.

\textbf{c.  Equivalent.}  Two geometric groups are equivalent ($\equiv$) if they have the same canonical listing of elements, except for the labelling of the element subscripts.  (The listing of a group is said to be canonical if all elements of the group are expressed as products of the vector generators of the Clifford group $\textbf{\textsf C}_{n,0}=\textbf{\textsf E}_{a_1}\textbf{\textsf E}_{a_2}\cdots\textbf{\textsf E}_{a_n}$.)  For example,
\begin{eqnarray}
\label{eq:E_a E_abc example}
\textbf{\textsf E}_{a}\textbf{\textsf E}_{abc}&=&\{\pm 1\}\{1,\mathbf e_a,\hat{\mathbf e}_{bc},\hat{\mathbf e}_{abc}\},\\
\label{eq:E_ab E_abc example}
\textbf{\textsf E}_{ab}\textbf{\textsf E}_{abc}&=&\{\pm 1\}\{1,\mathbf e_c,\hat{\mathbf e}_{ab},\hat{\mathbf e}_{abc}\}.
\end{eqnarray}
If we map the ordered pair $(a,c):\rightarrow (c,a)$, then Eq.~(\ref{eq:E_a E_abc}) becomes
\begin{eqnarray}
\label{eq:E_a E_abc c to a}
\textbf{\textsf E}_a\textbf{\textsf E}_{abc}:\rightarrow \textbf{\textsf E}_c\textbf{\textsf E}_{cba}&=&\{\pm 1\}\{1,\mathbf e_c,\hat{\mathbf e}_{ab},\hat{\mathbf e}_{abc}\}\nonumber\\
&=&\textbf{\textsf E}_{ab}\textbf{\textsf E}_{abc}.
\end{eqnarray}
Thus, 
\begin{equation}
\label{eq:E_a E_abc equiv E_ab E_abc}
\textbf{\textsf E}_a\textbf{\textsf E}_{abc}\equiv\textbf{\textsf E}_{ab}\textbf{\textsf E}_{abc}.
\end{equation}

\textbf{d.  Equal.} Two geometric groups are equal ($=$) if they have the same canonical listing of elements.  For example,
\begin{eqnarray}
\label{eq:E_a E_b is E_a E_ab}
\textbf{\textsf E}_a\textbf{\textsf E}_b&=&\{\pm 1\}\{1,\mathbf e_a\}\{1,\mathbf e_b\}\nonumber\\
&=&\{\pm 1\}\{1,\mathbf e_a,\mathbf e_b,\hat{\mathbf e}_{ab}\}\nonumber\\
&=&\{\pm 1\}\{1,\mathbf e_a\}\{1,\hat{\mathbf e}_{ab}\}=\textbf{\textsf E}_a\textbf{\textsf E}_{ab}.
\end{eqnarray}
Note that all equivalent groups can be made equal after relabelling the subscripts.  

Note also that if two single-generator groups are similar, then they are equal.  Thus, we may write Eq.~(\ref{eq:E_abc permute}) as a strict equality:
\begin{equation}
\label{eq:E_abc permute equality}
\textbf{\textsf E}_{abc}=\textbf{\textsf E}_{acb}=\textbf{\textsf E}_{bac}.
\end{equation}
This is obvious from the definition of $\textbf{\textsf E}_{abc}$ in Eq.~(\ref{eq:E_abc definition}).

\subsection{Hierarchy of Subgroups of $\textbf{\textsf C}_{3,0}$}

Let us enumerate the subgroups of $\textbf{\textsf C}_{3,0}$ according to the number of their nontrivial generators.  We shall limit ourselves to at most $n=3$ generators---though the maximum is $n=7$---to simplify the tables.  (Also, no Clifford basis subgroups can be formed for $\textbf{\textsf C}_{n,0}$ if the number of generators of the subgroup exceed $n$.)

\bigskip

\textbf{a.  One Generator.}  There are four $\textbf{\textsf C}_{3,0}$ subgroups with one nontrivial generator:
\begin{eqnarray}
\label{eq:E_a cong C_10}
\textbf{\textsf E}_a&=&\{\pm 1\}\{1,\mathbf e_a\}\cong\textbf{\textsf C}_{1,0},\\
\label{eq:E_ab cong C_01}
\textbf{\textsf E}_{ab}&=&\{\pm 1\}\{1,\hat{\mathbf e}_{ab}\}\cong\textbf{\textsf C}_{0,1},\\
\label{eq:E_abc cong C_01}
\textbf{\textsf E}_{abc}&=&\{\pm 1\}\{1,\hat{\mathbf e}_{abc}\}\cong\textbf{\textsf C}_{0,1}.
\end{eqnarray}
Thus, the Clifford group $\textbf{\textsf C}_{1,0}$ may be generated by a vector $\mathbf e_a$; the Clifford group $\textbf{\textsf C}_{0,1}$, by a bivector $\hat{\mathbf e}_{ab}$ or a trivector $\hat{\mathbf e}_{abc}$.  These groups are order four, with two linearly independent basis elements each.

Removing the trivial group 
\begin{equation}
\label{eq:C_00}
\textbf{\textsf C}_{0,0}=\{\pm 1\}=\{1,-1\}
\end{equation}
from Eq.~(\ref{eq:E_a cong C_10}), we obtain 
\begin{equation}
\{1,\mathbf e_a\}\cong\textbf{\textsf C}_{0,0}.
\end{equation}
Note that $\textbf{\textsf C}_{0,0}=\{\pm 1\}$ is the group basis for the algebra of $\textbf{\textsf R}$ of real numbers.

\textbf{b.  Two Generators.}  For geometric groups with two nontrivial generators, we shall order them by writing the product of single generator groups as $\textbf{\textsf E}_A\textbf{\textsf E}_B$ for $|A|\leq|B|$, where $A$ and $B$ are the multivector ranks of the generators.  For example, $\textbf{\textsf E}_{ab}$ is rank 2 (bivector generated group), while $\textbf{\textsf E}_{abc}$ is rank 3 (trivector generated group).  Thus, we write $\textbf{\textsf E}_{ab}\textbf{\textsf E}_{abc}$ and not $\textbf{\textsf E}_{abc}\textbf{\textsf E}_{ab}$.  Note that the subscripts could not be equal, $A\neq B$, because
\begin{equation}
\label{eq:E_A E_B is E_A E_A is E_A}
\textbf{\textsf E}_A\textbf{\textsf E}_B=\textbf{\textsf E}_A\textbf{\textsf E}_A=\textbf{\textsf E}_A,
\end{equation}
which is a group with only one nontrivial generator.

There are six $\textbf{\textsf C}_{3,0}$ subgroups with two nontrivial generators:
\begin{eqnarray}
\label{eq:E_a E_b}
\textbf{\textsf E}_a\textbf{\textsf E}_b&=&\{\pm 1\}\{1,\mathbf e_a,\mathbf e_b,\hat{\mathbf e}_{ab}\},\\
\label{eq:E_a E_ab}
\textbf{\textsf E}_a\textbf{\textsf E}_{ab}&=&\{\pm 1\}\{1,\mathbf e_a,\mathbf e_b,\hat{\mathbf e}_{ab}\}=\textbf{\textsf E}_a\textbf{\textsf E}_b,\\
\label{eq:E_a E_bc}
\textbf{\textsf E}_a\textbf{\textsf E}_{bc}&=&\{\pm 1\}\{1,\mathbf e_a,\hat{\mathbf e}_{bc},\hat{\mathbf e}_{abc}\},\\
\label{eq:E_a E_abc}
\textbf{\textsf E}_a\textbf{\textsf E}_{abc}&=&\textbf{\textsf E}_a\textbf{\textsf E}_{bc},\\
\label{eq:E_ab E_ac}
\textbf{\textsf E}_{ab}\textbf{\textsf E}_{ac}&=&\{\pm 1\}\{1,\hat{\mathbf e}_{ab},\hat{\mathbf e}_{ac},\hat{\mathbf e}_{bc}\},\\
\label{eq:E_ab E_abc}
\textbf{\textsf E}_{ab}\textbf{\textsf E}_{abc}&=&\{\pm 1\}\{1,\mathbf e_c,\hat{\mathbf e}_{ab},\hat{\mathbf e}_{abc}\}\nonumber\\
&\equiv&\textbf{\textsf E}_{bc}\textbf{\textsf E}_{abc}=\textbf{\textsf E}_a\textbf{\textsf E}_{bc}.
\end{eqnarray}
Notice that $\textbf{\textsf E}_a\textbf{\textsf E}_b$ and $\textbf{\textsf E}_a\textbf{\textsf E}_{ab}$ are canonically equal groups.  And so are the groups $\textbf{\textsf E}_a\textbf{\textsf E}_{bc}$, $\textbf{\textsf E}_a\textbf{\textsf E}_{abc}$, and $\textbf{\textsf E}_{ab}\textbf{\textsf E}_{abc}$, except that they are not isomorphic to Clifford basis groups, because their respective generators commute.  

Notice, too, that the set $\textbf{\textsf E}_{ab}\textbf{\textsf E}_{ac}$ is the set of quaternion basis:
\begin{eqnarray}
\label{eq:i quaternion}
\hat\imath&=&\hat{\mathbf e}_{ab},\\
\label{eq:j quaternion}
\hat\jmath&=&\hat{\mathbf e}_{bc},\\
\label{eq:k quaternion}
\hat k&=&\hat{\mathbf e}_{ac}.
\end{eqnarray}

\textbf{c.  Three Generators.}  There are ten subgroups with three nontrivial generators:
\begin{eqnarray}
\label{eq:E_a E_b E_c}
\textbf{\textsf E}_a\textbf{\textsf E}_b\textbf{\textsf E}_c&=&\{\pm 1\}\{1,\mathbf e_a,\mathbf e_b,\mathbf e_c,\nonumber\\
& &\qquad\quad\hat{\mathbf e}_{ab},\hat{\mathbf e}_{ac},\hat{\mathbf e}_{bc},\hat{\mathbf e}_{abc}\},\\
\label{eq:E_a E_b E_ab}
\textbf{\textsf E}_a\textbf{\textsf E}_b\textbf{\textsf E}_{ab}&=&\textbf{\textsf E}_a\textbf{\textsf E}_b,\\
\label{eq:E_a E_b E_ac}
\textbf{\textsf E}_a\textbf{\textsf E}_b\textbf{\textsf E}_{ac}&=&\textbf{\textsf E}_a\textbf{\textsf E}_b\textbf{\textsf E}_c,\\
\label{eq:E_a E_b E_abc}
\textbf{\textsf E}_a\textbf{\textsf E}_b\textbf{\textsf E}_{abc}&=&\textbf{\textsf E}_a\textbf{\textsf E}_b\textbf{\textsf E}_c,\\
\label{eq:E_a E_ab E_ac}
\textbf{\textsf E}_a\textbf{\textsf E}_{ab}\textbf{\textsf E}_{ac}&=&\textbf{\textsf E}_a\textbf{\textsf E}_b\textbf{\textsf E}_c,\\
\label{eq:E_a E_ab E_bc}
\textbf{\textsf E}_a\textbf{\textsf E}_{ab}\textbf{\textsf E}_{bc}&=&\textbf{\textsf E}_a\textbf{\textsf E}_b\textbf{\textsf E}_c,\\
\label{eq:E_a E_ab E_abc}
\textbf{\textsf E}_a\textbf{\textsf E}_{ab}\textbf{\textsf E}_{abc}&=&\textbf{\textsf E}_a\textbf{\textsf E}_b\textbf{\textsf E}_c,\\
\label{eq:E_a E_bc E_abc}
\textbf{\textsf E}_a\textbf{\textsf E}_{bc}\textbf{\textsf E}_{abc}&=&\textbf{\textsf E}_a\textbf{\textsf E}_{bc},\\
\label{eq:E_ab E_ac E_bc}
\textbf{\textsf E}_{ab}\textbf{\textsf E}_{ac}\textbf{\textsf E}_{bc}&=&\textbf{\textsf E}_{ab}\textbf{\textsf E}_{ac},\\
\label{eq:E_ab E_ac E_abc}
\textbf{\textsf E}_{ab}\textbf{\textsf E}_{ac}\textbf{\textsf E}_{abc}&=&\textbf{\textsf E}_a\textbf{\textsf E}_b\textbf{\textsf E}_c,
\end{eqnarray}
Notice that no two of these groups are isomorphic, though some of them are equal.  Notice, too, that though $\textbf{\textsf E}_{ab}\textbf{\textsf E}_{ac}\textbf{\textsf E}_{bc}$ is canonically equal to the quaternion group $\textbf{\textsf E}_{ab}\textbf{\textsf E}_{ac}$, they are not isomorphic: the latter is isomorphic to $\textbf{\textsf C}_{0,2}$, but the former tries to be isomorphic to $\textbf{\textsf C}_{0,3}$.

\subsection{Choirs and Bands}

Let us classify the subgroups of $\textbf{\textsf C}_{3,0}$ according their obedience to the three rules for generating \emph{universal} Clifford algebras\cite{Porteous_2000pp126-130}:
\begin{itemize}
\item The square of each generator is $+1$ or $-1$
\item All nontrivial generators in the same group anticommute
\item The order (number of elements) of the group is $2^{n+1}$ (+ and $-$ elements are distinct), where $n$ is the group's number of nontrivial generators.
\end{itemize}
Obedient groups we shall call \emph{choirs}, in analogy to the angelic hierarchy devised by Dionysius the Pseudo-Areopagite\cite{Stiglmayr_1909}\cite{Parker_1897}.   Disobedient groups we shall call \emph{bands}. 
\bigskip

\textbf{a.  Choirs.}  There are two ways to arrange the choirs.  One way is to arrange them according to the signature of the Clifford algebras they form (see Table 1).  Another way is to arrange them according to \emph{modes} (in analogy to the eightfold way of classifying Gregorian chants\cite{Meyer_1952}): groups with the same mode are canonically equal ($=$).  We shall designate the foremost choir in each mode as the leader or \emph{cantor}.  (See Tables 2 to 5)

Notice that there are nine choirs in $\textbf{\textsf C}_{3,0}$ as in the Dionysian system.  The groupings, however, do not follow Dionysius's pattern of 3-3-3 but of 1-3-3-2.  Notice, too, that there are four chant modes, with the mode number corresponding to the number of generators.

\textbf{b.  Bands}  There are two ways to arrange the bands.  One way is to arrange them according to signature of the Clifford algebra they failed to obey (see Table 6).  Another way is to arrange them according to \emph{rhythms}: bands in the same rhythm are canonically equal ($=$).  We shall designate in each rhythm a \emph{leader}.  (See Tables 7 to 10)

Three useful measures for describing bands are disorder, chord, and beat.  

The \emph{disorder} $\Phi$ of a group of order $2^m$ with $n$ nontrivial generators is
\begin{equation}
\label{eq:disorder delta_mn}
\Phi=\log_2\frac{2^{n+1}}{2^m}=n+1-m.
\end{equation}
Notice that disorder is measured with respect to $2^{n+1}$, which is the order prescribed for choirs.  Choirs have a disorder $\Phi=0$.

The number of generators in the same group that the generator commutes with (going into another's place without making a $-$ sign) other than itself is called the generator's transposition number or simply \emph{transposition}.  The series of transpositions of each generator describes the band's chord $X$.  

The sum $T$ of the transpositions of the band's generators divided by the total possible permutations of $n$ of generators taken $2$ at a time is the band's \emph{beat}:
\begin{equation}
\label{eq:malice B}
B=T\frac{(n-2)!}{n!}=\frac{T}{n(n-1)}.
\end{equation}
Bands have a maximum beat of $B=1$.  Choirs have a beat of $B=0$.  Note that beat for sigle-generator groups is not defined. 

Because we associated choir groups with the angelic hierarchy, let us, for mnemonic purposes, associate band groups with the demonic lowerarchy, as suggested by their dominant number 6-6-6.  Beatless bands with a unity disorder are $\{1,\mathbf e_a\}$, $\textbf{\textsf E}_a\textbf{\textsf E}_b\textbf{\textsf E}_{ab}$, and $\textbf{\textsf E}_{ab}\textbf{\textsf E}_{ac}\textbf{\textsf E}_{bc}$.  Most bands have no unity disorder but have nonzero beats.  One band stands out with a disorder of one and a beat of one: $\textbf{\textsf E}_{a}\textbf{\textsf E}_{bc}\textbf{\textsf E}_{abc}$.

There are still many demonic bands whose number of generators fall in the interval $4\leq n\leq 7$.  They are legion.

\section{Conclusion}

In this paper, we showed that the basis set of the Clifford $\mathcal Cl_{n,0}$ algebra forms a group under juxtaposition multiplication.  This group with $2^{n+1}$ elements is generated by $n$ anticommuting vectors that square to $+1$.  Using set algebra, we showed that from this basis set we can construct geometric groups whose generators not only square to $\pm 1$, but also commute or anticommute with other generators in the group.  

To illustrate this claim, we enumerated all the subgroups of the basis set of $\mathcal Cl_{3,0}$ according to the number $n$ of their nontrivial generators, for $n\leq 3$.  We classified the subgroups according to three criteria: (1) the square of each generator is $\pm 1$, (2) the generators within the group anticommute, and (3) the order of the resulting group is $2^{n+1}$.  All obey the first rule, but some groups fail in the second or third rule or both.  Obedient groups we called choirs; disobedient groups, bands.  

The choirs form the basis of Clifford algebras of arbitrary signature for $p+q=n$.  Canonically equal choirs we grouped into modes.  Bands, on the other hand, do not form the basis of Clifford algebras because of their disobedience.  We distinguished bands according to disorder, chord, and beat.  Canonically equal bands we classified under one rhythm.

A similar taxonomic system may be made for the basis set of the Clifford (Dirac) algebra $\mathcal Cl_{4,0}$.  But this may be a Herculean task, requiring days and weeks of pen and paper computations.  A computer may be necessary.

\section*{\small{Acknowledgments}}
This research was supported by the Manila Observatory and by the Physics Department of Ateneo de Manila University.

\newpage

\begin{table}
\footnotesize
\begin{center}
\begin{tabular}{|l|l|l|c|l|}\hline
$n$\T & $\cong$ &\textbf{Choir} & \textbf{Sign} & Name\\\hline\hline
0 \T& $\textbf{\textsf C}_{0,0}$ & $\{\pm 1\}$ & + & Seraphim\\\hline\hline
1 \T& $\textbf{\textsf C}_{1,0}$ & $\textbf{\textsf E}_a$& + & Cherubim\\\hline
1 \T& $\textbf{\textsf C}_{0,1}$ & $\textbf{\textsf E}_{ab}$ & $-$ & Thrones\\\hline 
1 \T& $\textbf{\textsf C}_{0,1}$ & $\textbf{\textsf E}_{abc}$ & $-$ & Virtues\\\hline \hline
2 \T& $\textbf{\textsf C}_{2,0}$ & $\textbf{\textsf E}_a\textbf{\textsf E}_b$ & $++$ & Dominations\\\hline 
2 \T& $\textbf{\textsf C}_{1,1}$ & $\textbf{\textsf E}_a\textbf{\textsf E}_{ab}$ & $+-$ & Powers\\\hline
2 \T& $\textbf{\textsf C}_{0,2}$ & $\textbf{\textsf E}_{ab}\textbf{\textsf E}_{ac}$ & $--$ & Principalities\\\hline \hline
3 \T& $\textbf{\textsf C}_{3,0}$ & $\textbf{\textsf E}_a\textbf{\textsf E}_b\textbf{\textsf E}_c$ & $+++$ & Archangels\\\hline 
3 \T& $\textbf{\textsf C}_{1,2}$ & $\textbf{\textsf E}_a\textbf{\textsf E}_{ab}\textbf{\textsf E}_{ac}$ & $+--$ & Angels\\\hline 
\end{tabular}
\end{center}
\caption{\footnotesize Hierarchy of choir groups and their angelic names}
\end{table}

\begin{table}
\footnotesize
\begin{center}
\begin{tabular}{|l|l|l|c|}\hline
$n$\T & $\cong$ &\textbf{Choir} & \textbf{Sign}\\\hline
0 \T& $\textbf{\textsf C}_{0,0}$ & $\{\pm 1\}$ & +\\\hline
\end{tabular}
\end{center}
\caption{\footnotesize Mode 0 by $\{\pm 1\}$}
\end{table}

\begin{table}
\footnotesize
\begin{center}
\begin{tabular}{|l|l|l|c|}\hline
$n$\T & $\cong$ &\textbf{Choir} & \textbf{Sign}\\\hline
1 \T& $\textbf{\textsf C}_{1,0}$ & $\textbf{\textsf E}_a$& +\\\hline
2 \T& $\textbf{\textsf C}_{0,1}$ & $\textbf{\textsf E}_{ab}$ & $-$\\\hline
3 \T& $\textbf{\textsf C}_{0,1}$ & $\textbf{\textsf E}_{abc}$ & $-$\\\hline
\end{tabular}
\end{center}
\caption{\footnotesize Mode 1 led by $\textbf{\textsf E}_a$}
\end{table}

\begin{table}
\footnotesize
\begin{center}
\begin{tabular}{|l|l|l|c|}\hline
$n$\T & $\cong$ &\textbf{Choir} & \textbf{Sign}\\\hline\hline
2 \T& $\textbf{\textsf C}_{2,0}$ & $\textbf{\textsf E}_a\textbf{\textsf E}_b$ & $++$\\\hline
2 \T& $\textbf{\textsf C}_{1,1}$ & $\textbf{\textsf E}_a\textbf{\textsf E}_{ab}$& $+-$\\\hline
2 \T& $\textbf{\textsf C}_{0,2}$ & $\textbf{\textsf E}_{ab}\textbf{\textsf E}_{ac}$& $--$\\\hline
\end{tabular}
\end{center}
\caption{\footnotesize Mode 2 led by $\textbf{\textsf E}_a\textbf{\textsf E}_b$}
\end{table}

\begin{table}
\footnotesize
\begin{center}
\begin{tabular}{|l|l|l|c|}\hline
$n$\T & $\cong$ &\textbf{Choir} & \textbf{Sign}\\\hline\hline
3 \T& $\textbf{\textsf C}_{3,0}$ & $\textbf{\textsf E}_a\textbf{\textsf E}_b\textbf{\textsf E}_c$ & $+++$\\\hline
3 \T& $\textbf{\textsf C}_{1,2}$ & $\textbf{\textsf E}_{a}\textbf{\textsf E}_{ab}\textbf{\textsf E}_{ac}$ & $+--$\\\hline
\end{tabular}
\end{center}
\caption{\footnotesize Mode 3 led by $\textbf{\textsf E}_a\textbf{\textsf E}_b\textbf{\textsf E}_c$}
\end{table}

\begin{table}
\footnotesize
\begin{center}
\begin{tabular}{|l|l|l|c|l|l|l|}
\hline
$n$\T & $\cong\!\!\!\!\!\!/$ &\textbf{Band} & \textbf{Sign} & $\Phi$ & $\quad\ \ X$ & $B$\\
\hline\hline
1 \T& $\textbf{\textsf C}_{1,0}$ & $\{1,\mathbf e_a\}$ & $+$ & 1 & & \\
\hline
2 \T& $\textbf{\textsf C}_{1,1}$ & $\textbf{\textsf E}_a\textbf{\textsf E}_{bc}$ & $+-$ & 0 & $(1,1)$ & 2/2 \\
\hline
2 \T& $\textbf{\textsf C}_{1,1}$ & $\textbf{\textsf E}_a\textbf{\textsf E}_{abc}$ & $+-$ & 0 & $(1,1)$ & 2/2 \\
\hline
2 \T& $\textbf{\textsf C}_{0,2}$ & $\textbf{\textsf E}_{ab}\textbf{\textsf E}_{abc}$ & $--$ & 0 & $(1,1)$ & 2/2 \\
\hline
3 \T& $\textbf{\textsf C}_{2,1}$ & $\textbf{\textsf E}_{a}\textbf{\textsf E}_{b}\textbf{\textsf E}_{ab}$ & $++-$ & 1 & $(0,0,0)$ & 0/6 \\
\hline
3 \T& $\textbf{\textsf C}_{2,1}$ & $\textbf{\textsf E}_{a}\textbf{\textsf E}_{b}\textbf{\textsf E}_{ac}$ & $++-$ & 0 & $(0,1,1)$ & 2/6 \\
\hline
3 \T& $\textbf{\textsf C}_{2,1}$ & $\textbf{\textsf E}_{a}\textbf{\textsf E}_{b}\textbf{\textsf E}_{abc}$ & $++-$ & 0 & $(1,1,2)$ & 4/6 \\
\hline
3 \T& $\textbf{\textsf C}_{1,2}$ & $\textbf{\textsf E}_{a}\textbf{\textsf E}_{ab}\textbf{\textsf E}_{bc}$ & $+--$ & 0 & $(1,0,1)$ & 2/6 \\
\hline
3 \T& $\textbf{\textsf C}_{1,2}$ & $\textbf{\textsf E}_{a}\textbf{\textsf E}_{ab}\textbf{\textsf E}_{abc}$ & $+--$ & 0 & $(1,1,2)$ & 4/6 \\
\hline
3 \T& $\textbf{\textsf C}_{1,2}$ & $\textbf{\textsf E}_{a}\textbf{\textsf E}_{bc}\textbf{\textsf E}_{abc}$ & $+--$ & 1 & $(2,2,2)$ & 6/6 \\
\hline
3 \T& $\textbf{\textsf C}_{0,3}$ & $\textbf{\textsf E}_{ab}\textbf{\textsf E}_{ac}\textbf{\textsf E}_{bc}$ & $---$ & 1 & $(0,0,0)$ & 0/6 \\
\hline
3 \T& $\textbf{\textsf C}_{0,3}$ & $\textbf{\textsf E}_{ab}\textbf{\textsf E}_{ac}\textbf{\textsf E}_{abc}$ & $---$ & 0 & $(1,1,2)$ & 4/6 \\
\hline
\end{tabular}
\end{center}
\caption{\footnotesize Lowerarchy of band groups arranged according to signature $(p,q)$, disorder $\Phi$, chord $X$, and beat $B$}
\end{table} 

\begin{table}
\footnotesize
\begin{center}
\begin{tabular}{|l|l|l|c|l|l|l|}
\hline
$n$\T & $\cong\!\!\!\!\!\!/$ &\textbf{Band} & \textbf{Sign} & $\Phi$ & $\Gamma$ & $B$\\
\hline\hline
1 \T& $\textbf{\textsf C}_{1,0}$ & $\{1,\mathbf e_a\}$ & $+$ & $1$ & & \\
\hline
\end{tabular}
\end{center}
\footnotesize
\caption{\footnotesize The band $\{1,\mathbf e_a\}$ with disorder $\Phi=1$, undefined chord $X$, and undefined beat $B$}
\end{table} 

\begin{table}
\footnotesize
\begin{center}
\begin{tabular}{|l|l|l|c|l|l|l|}
\hline
$n$\T & $\cong\!\!\!\!\!\!/$ &\textbf{Band} & \textbf{Sign} & $\Phi$ & $X$ & $B$\\
\hline\hline
2 \T& $\textbf{\textsf C}_{1,1}$ & $\textbf{\textsf E}_a\textbf{\textsf E}_{bc}$ & $+-$ & 0 & $(1,1)$ & 2/2 \\
\hline
2 \T& $\textbf{\textsf C}_{1,1}$ & $\textbf{\textsf E}_a\textbf{\textsf E}_{abc}$ & $+-$ & 0 & $(1,1)$ & 2/2 \\
\hline
2 \T& $\textbf{\textsf C}_{0,2}$ & $\textbf{\textsf E}_{bc}\textbf{\textsf E}_{abc}$ & $--$ & 0 & $(1,1)$ & 2/2 \\
3 \T& $\textbf{\textsf C}_{1,2}$ & $\textbf{\textsf E}_{a}\textbf{\textsf E}_{bc}\textbf{\textsf E}_{abc}$ & $+--$ & 1 & $(2,2,2)$ & 6/6 \\
\hline
\end{tabular}
\end{center}
\caption{\footnotesize Rhythm of $\textbf{\textsf E}_{a}\textbf{\textsf E}_{bc}$}
\end{table} 

\begin{table}
\footnotesize
\begin{center}
\begin{tabular}{|l|l|l|c|l|l|l|}
\hline
$n$\T & $\cong\!\!\!\!\!\!/$ &\textbf{Band} & \textbf{Sign} & $\Phi$ & $X$ & $B$\\
\hline\hline
3 \T& $\textbf{\textsf C}_{2,1}$ & $\textbf{\textsf E}_{a}\textbf{\textsf E}_{b}\textbf{\textsf E}_{ab}$ & $++-$ & 1 & $(0,0,0)$ & 0/6\\
\hline
\end{tabular}
\end{center}
\caption{\footnotesize Rhythm of $\textbf{\textsf E}_{a}\textbf{\textsf E}_{b}\textbf{\textsf E}_{ab}$}
\end{table} 

\begin{table}[t]
\footnotesize
\begin{center}
\begin{tabular}{|l|l|l|c|l|l|l|}
\hline
$n$\T & $\cong\!\!\!\!\!\!/$ &\textbf{Band} & \textbf{Sign} & $\Phi$ & $X$ & $B$\\
\hline\hline
3 \T& $\textbf{\textsf C}_{2,1}$ & $\textbf{\textsf E}_{a}\textbf{\textsf E}_{b}\textbf{\textsf E}_{ac}$ & $++-$ & 0 & $(0,1,1)$ & 2/6 \\
\hline
3 \T& $\textbf{\textsf C}_{2,1}$ & $\textbf{\textsf E}_{a}\textbf{\textsf E}_{b}\textbf{\textsf E}_{abc}$ & $++-$ & 0 & $(1,1,2)$ & 4/6 \\
\hline
3 \T& $\textbf{\textsf C}_{1,2}$ & $\textbf{\textsf E}_{a}\textbf{\textsf E}_{ab}\textbf{\textsf E}_{bc}$ & $+--$ & 0 & $(1,0,1)$ & 2/6 \\
\hline
3 \T& $\textbf{\textsf C}_{1,2}$ & $\textbf{\textsf E}_{a}\textbf{\textsf E}_{ab}\textbf{\textsf E}_{abc}$ & $+--$ & 0 & $(1,1,2)$ & 4/6 \\
\hline
3 \T& $\textbf{\textsf C}_{0,3}$ & $\textbf{\textsf E}_{ab}\textbf{\textsf E}_{ac}\textbf{\textsf E}_{abc}$ & $---$ & 0 & $(1,1,2)$ & 4/6 \\
\hline
\end{tabular}
\end{center}
\caption{\footnotesize Rhythm of $\textbf{\textsf E}_{a}\textbf{\textsf E}_{b}\textbf{\textsf E}_{ac}$}
\end{table}

\end{document}